\documentclass[aps,prl,twocolumn, groupaddress]{revtex4}
\usepackage{graphicx}
\usepackage{dcolumn}
\usepackage{amsmath}
\usepackage{amssymb}
\usepackage{subfigure,amsmath,verbatim,moreverb,bm}
\def\be{\begin{equation}}
\def\ee{\end{equation}}
\def\ber{\begin{eqnarray}}
\def\eer{\end{eqnarray}}

\def\nablabold{\mbox{\boldmath $\nabla$}}
\def\rv{{\bf r}}

\def\jv{{\bf j}}

\def\Av{{\bf A}}

\def\uv{{\bf u}}

\def\fv{\bf{\cal F}}
\def\nn{\nonumber}

\begin{document}
\title{Linear Continuum Mechanics for Quantum Many-Body Systems}
\author{Jianmin Tao$^1$} 
\author{Xianlong Gao$^{2}$} 
\thanks{Permanent address: Department of Physics, Zhejiang Normal University, Jinhua, Zhejiang Province, 321004, China}
\author{G. Vignale$^2$}
\author{I. V. Tokatly$^{3,4}$}
\affiliation{$^1$Theoretical Division and CNLS,
Los Alamos National Laboratory, Los Alamos, New Mexico 87545\\$^2$Department of Physics, University of Missouri-Columbia,
Columbia, Missouri 65211\\$^3$European Theoretical Spectroscopy Facility (ETSF), Departamento de Fisica de Materiales, \\ Universidad del Pais Vasco UPV/EHU, Centro Mixto CSIC-UPV/EHU, E-20018 San Sebasti\'an,  Spain\\
$^4$Moscow Institute of Electronic Technology, Zelenograd, 124498 Russia}
\date{\today}
\begin{abstract}
We develop the continuum mechanics of quantum many-body systems in the linear response regime.  The basic variable of the theory is the displacement field,  for which we derive a closed equation of motion under the assumption that the time-dependent wave function in a locally co-moving reference frame can be described  as a geometric deformation of the ground-state wave function.   We show that this equation of motion is exact for systems consisting of a single particle, and for all systems at sufficiently high frequency, and that it leads to an excitation spectrum that has the correct integrated strength. The theory is illustrated by simple model applications to one- and  two-electron systems.
\par
\end{abstract}
\maketitle
The dynamics of quantum many-particle systems, as displayed in electromagnetic transitions, chemical reactions, ionization and collision processes, poses a major challenge to computational physicists and chemists.   Whereas the calculation of ground-state properties can be tackled by powerful computational methods such as the quantum Monte Carlo,~\cite{MonteCarlo} 
the development of similar methods for time-dependent properties has been slow.  One of the most successful methods to date  is the time-dependent density functional theory (TDDFT), or its more recent version -- time-dependent {\it current} density functional theory (TDCDFT).\cite{grossbook}   In the common Kohn-Sham implementation of this method\cite{Gross96,Casida95}  the formidable problem of solving the time-dependent Schr\"odinger equation for the many-body wave function is replaced by the much simpler problem of determining $N$ single-particle orbitals.   However, even this simplified problem is quite complex,  and furthermore there are features such as multi-particle excitations \cite{Maitra04} and dispersion forces \cite{Dobson98}  that are very difficult to treat within the conventional approximation schemes.   

An alternative approach, which actually dates back to the early days of quantum mechanics \cite{TerHaar,Madelung1927,Bloch1933},  attempts to calculate directly the collective variables of interest -- density and current.  This approach we call ``quantum continuum mechanics" (QCM), because in analogy with classical theories of continuous media  it attempts to describe the quantum many-body system without explicit reference to the individual particles.\cite{GhoDeb1982}


The possibility of a QCM formulation of the quantum many-body problem is guaranteed by the very same theorems that lay down the foundation of TDDFT and TDCDFT.\cite{rg,vanLeeuwen99}   Let us consider a system of particles described by the time-dependent Hamiltonian
\be\label{defH}
\hat H(t) = \hat H_0+\int d\rv \hat n(\rv)V_1(\rv,t)
\ee
where  
$
\hat H_0 = \hat T +\hat W+\hat V_0
$
is the sum of kinetic energy ($\hat T$),  interaction potential energy ($\hat W$),  and the potential energy associated with an external {\it static} potential ($\hat V_0$).  $\hat n(\rv)$ is the particle density operator and   $V_1(\rv,t)$  is an external time-dependent potential.  The exact Heisenberg equation of motion for  the current density operator, averaged over the quantum state, leads to the Euler equation
\ber\label{ForceBalance} 
m\partial_t j_\mu(\rv,t)=
&-&n(\rv,t) \partial_\mu [V_0(\rv)+V_1(\rv,t)]\nn\\
&-&\partial_\nu P_{\mu\nu}(\rv,t)\,. 
\eer
Here $m$ is the mass of the particles and repeated indices are summed over.   The key quantity on the right hand side of Eq.~(\ref{ForceBalance}) is the {\it stress tensor} $P_{\mu\nu}(\rv,t)$ --  a symmetric tensor whose divergence yields the {\it force density} arising from quantum-kinetic and interaction effects.  Now the Runge-Gross theorem of TDDFT guarantees that the stress tensor, like every observable of the system, is a functional of the current density  and of the initial quantum state.  Thus,  Eq.~(\ref{ForceBalance}) is in principle a closed equation of motion for $\jv$ -- the only missing piece
being the {\it explicit} expression for $P_{\mu\nu}$ in terms of the
current density.   

In recent years much effort has been devoted to the theoretical problem of constructing an approximate QCM\cite{Zaremba94,ConVig1999,TokPanPRB,DobLe,tokatly,tvt07} and several applications have appeared in the literature (see Ref.~\onlinecite{Applications} for some representative examples).   All approximation schemes so far have been based on the local density approximation and generalizations thereof.  In this Letter we derive a new approximate expression for
$P_{\mu\nu}(\rv,t)$ as a functional of the current density for systems that perform small amplitude oscillations about the ground-state.  The new formula is nonlocal, is expressed in terms of calculable ground-state properties, and becomes exact in the high-frequency limit. 

The Euler equation~(\ref{ForceBalance})  is conveniently expressed in terms of the displacement field $\uv(\rv,t)$, which in the linear  regime is defined by the relation
$\jv(\rv,t) = n_0(\rv) \partial_t \uv(\rv,t)$,
where $n_0(\rv)$ is the ground-state density.  It is also convenient to write the density and the stress tensor as the sum of a large ground-state component and a small time-dependent part, i.e., $n(\rv,t)=n_0(\rv)+n_1(\rv,t)$ and  $P_{\mu\nu}(\rv,t)=P_{\mu\nu,0}(\rv)+P_{\mu\nu,1}(\rv,t)$. 
Then the time-dependent components satisfy the linearized form of the 
Euler equation ~(\ref{ForceBalance}) 
\be\label{eom.1}
m n_0(\rv)\partial_t^2 \uv=-n_0(\rv)\nablabold V_1(\rv,t) +\fv_1(\rv,t)\,,
\ee
where the total force density
\be\label{DefFv1}
{\cal F}_{\mu,1}(\rv,t) \equiv -n_1(\rv,t)\partial_\mu V_0(\rv)-\partial_\nu P_{\mu\nu,1}(\rv,t)\,,
\ee 
is  a linear functional of $\uv(\rv,t)$.  
Our approximate expression for $\fv_{\mu,1}$ will be presented in terms of the functional 
\be\label{defEu}
E[\uv] \equiv \langle \psi_0[\uv]|\hat H_0|\psi_0[\uv]\rangle\,,
\ee
which is the energy of the deformed ground-state $|\psi_0[\uv]\rangle$, obtained from the undistorted ground-state $|\psi_0\rangle$ by displacing the volume element  located at $\rv$ to a new position $\rv+\uv(\rv,t)$.  More precisely, we will argue that the force density can be represented as
\be\label{fv1}
\fv_{\mu,{\rm 1}}(\rv,t)=-\int d\rv' \left.\frac{\delta^2 {\it E}[\uv]}{\delta u_\mu (\rv)\delta u_\nu (\rv')}\right\vert_{\uv=0} u_\nu(\rv',t)\,,
\ee
where the second variational derivative of $E[\uv]$, evaluated at the ground-state ($\uv=0$)  has an exact expression in terms of the  one- and two-particle density matrices of the {\it ground-state}.   We will show that the representation~(\ref{fv1}) is exact for all one-particle systems and also for many-particle systems at sufficiently high frequency.  

Eq.~(\ref{fv1}) can be derived by performing a transformation to the ``co-moving reference frame"\cite{tokatly,tvt07} -- a non-inertial frame in which the density is constant and equal to the ground state density and the current density is zero -- and assuming that the wave function in this frame is independent of time.  This assumption is absolutely correct in one-particle systems, where the constancy of the density and the vanishing of the current density completely determine the wave function.  It is also generally valid on very short time scales, or for frequencies higher than the characteristic energy of single-particle excitations,  because on these time scales it is not possible for the particles to ``forget" the correlations built into the initial ground-state wave function.  
In all other cases our approximation replaces the exact ``normal modes" of the system by a smaller set of approximate normal modes, in such a way that the total spectral weight is conserved.   We now present a simple derivation of Eq.~(\ref{fv1}), which allows us to quickly recognize these facts. 

We start from the linear response of the current density to an external vector potential of frequency $\omega$
\be\label{LinearResponse1}
j_\mu(\rv,\omega) = \int d\rv' \chi_{\mu\nu}(\rv,\rv',\omega)A_{\nu,1}(\rv',\omega)\,,
\ee
where $j_\mu(\rv,\omega)$ is the Fourier component of the current at frequency $\omega$ and $\chi_{\mu\nu}(\rv,\rv',\omega)$ is the current-current response function.  At high frequency, $\chi_{\mu\nu}$  has the well-known expansion\cite{gvbook}
\be\label{LinearResponse2}
\chi_{\mu\nu}(\rv,\rv',\omega)=\frac{n_0(\rv)}{m}\delta(\rv-\rv')\delta_{\mu\nu}+\frac{M_{\mu\nu}(\rv,\rv')}{m^2\omega^2}\,,
\ee
where the first term (diamagnetic) is frequency-independent and
\ber
M_{\mu\nu}(\rv,\rv') \equiv -m^2 \langle \Psi_0 \vert [[\hat H_0,\hat j_\mu(\rv)],\hat j_\nu(\rv')]\vert\Psi_0\rangle\,,
\eer
is the first spectral moment of the current-current  response function.
Now, substituting  Eq.~(\ref{LinearResponse2})   in Eq.~(\ref{LinearResponse1})  and noting that
$\jv(\rv,\omega)=-i\omega n_0(\rv)\uv(\rv,\omega)$ and that the vector potential is related to the scalar potential by the equation $\Av_1(\rv,\omega) = \frac{\nablabold V_1(\rv,\omega)}{i\omega}$,  we obtain (to leading order in $1/\omega^2$): 
\be\label{eom.last}
-m \omega^2 n_0 u_\mu = -n_0\partial_\mu V_1 -\int d\rv' M_{\mu\nu}(\rv,\rv') u_\nu(\rv',\omega)\,.
\ee
This is equivalent to our equation of motion~(\ref{eom.1}), with $\fv_{\mu,1}$ given by Eq.~(\ref{fv1}),  if and only if 
\be\label{Moment-Energy}
 M_{\mu\nu}(\rv,\rv')= \left.\frac{\delta^2 E[\uv]}{\delta u_\mu(\rv)\delta u_\nu(\rv')}\right\vert_{\uv=0}\,.
 \ee
To show that this is the case we observe that the deformed ground-state is related to the undeformed ground-state by the unitary tranformation
\be\label{UnitaryTransformation}
|\Psi_0[\uv]\rangle = e^{-i\int d\rv \hat j(\rv)\cdot \uv(\rv)}|\Psi_0\rangle\,.
\ee
Here we have used the fact that the current density operator $\hat \jv(\rv)$ is the generator of a translation of all the particles in an infinitesimal volume located at $\rv$.  
Thus, the transformation (\ref{UnitaryTransformation})  amounts to performing different translations by vectors $\uv(\rv)$ at different points in space, i.e. precisely to deforming the system according to the displacement field $\uv(\rv)$.  
Substituting the above expression for $|\Psi_0[\uv]\rangle$ in the definition of $E[\uv]$ and expanding to second order in $\uv$ we can easily verify that
\be
E[\uv]\simeq E_0+\frac{1}{2}\int  d\rv \int d\rv' u_\mu(\rv) M_{\mu\nu}(\rv,\rv') u_\nu(\rv')\,,
\ee
which establishes the validity of Eq.~(\ref{Moment-Energy}).   

A lengthy calculation allows us to calculate the three components of the force density functional arising from the kinetic, interaction, and external potential parts of the Hamiltonian: $\fv_{\mu,{\rm 1}} = \fv^{\rm kin}_{\mu,{\rm 1}}+\fv^{\rm int}_{\mu,{\rm 1}}+\fv^{\rm pot}_{\mu,{\rm 1}}$.  The final results are
\ber\label{Fk}
&&\fv^{\rm kin}_{\mu,{\rm 1}} =
\partial_\alpha[{\rm 2}{\it T}_{\nu\mu,0}{\it u}_{\nu\alpha}+{\it T}_{\nu\alpha,0}\partial_\mu {\it u}_\nu]\nn\\
&-&\frac{1}{4m}\partial_\nu\partial_\mu (n_0\partial_\nu \nabla\cdot\uv)\nn\\
&+&\frac{1}{4m}\partial_\nu \left\{2(\nabla^2 n_0) u_{\nu\mu}+(\partial_\nu n_0)\partial_\mu \nabla\cdot\uv\right.\nn\\
&+&\left.(\partial_\mu n_0)\partial_\nu \nabla\cdot\uv-2\partial_\mu\left[ (\partial_\alpha n_0)u_{\nu\alpha}\right]\right\}\,,
\eer

\be\label{Fint}
\fv^{\rm int}_{\mu,{\rm 1}} = \int {\it d} \rv'  {\it K}_{\mu\nu}(\rv,\rv') [{\it u}_\nu(\rv)-{\it u}_\nu (\rv')]\,,
\ee
\be\label{Fpot}
\fv^{\rm pot}_{\mu,{\rm 1}} = -{\it n}_{\rm 0}(\rv)\uv\cdot\nablabold \partial_\mu {\it V}_{\rm 0}\,.
\ee

Here we have introduced the equilibrium stress tensor
\be\label{Tmunu0}
T_{\mu\nu,0}=\frac{1}{2m}\left(\partial_\mu\partial_\nu^\prime+\partial_\nu\partial_\mu^\prime\right) \rho^{(1)}(\rv,\rv')\Big\vert_{\rv=\rv'} - \frac{1}{4m}\nabla^2 n_0\delta_{\mu\nu}\,,
\ee
where $\rho^{(1)}(\rv,\rv')$ is the one-particle density matrix.   The interaction kernel $K$ in Eq.~(\ref{Fint}) is given by
\be\label{CoulombKernel}
K_{\mu\nu}(\rv,\rv')=\rho_2(\rv,\rv')\partial_\mu\partial_\nu^\prime w(|\rv-\rv'|)\,,
\ee
where $w(|\rv-\rv'|)$ is the interaction potential and $\rho_2(\rv,\rv')\equiv \rho^{(2)}(\rv,\rv'|\rv,\rv')$, where $\rho^{(2)}$  is the two-particle density matrix.  
Notice that the kinetic force density $\fv^{\rm kin}_{\mu,{\rm 1}}$ is a {\it semilocal} functional of the displacement, involving only derivatives up to the fourth order.

The excitation energies of the system are obtained from the solution  of  Eq.~(\ref{eom.last}) after setting $V_1=0$.  This equation defines a hermitian eigenvalue problem with positive eigenvalues $\omega_n^2$ -- the square of the excitation energies.  The positivity follows from the fact that a deformation of the ground-state wave function must necessarily increase the energy.   The corresponding eigenfunctions $\uv_n(\rv)$ are mutually orthogonal with respect to the scalar product  $(\uv_n,\uv_m) \equiv \int d\rv u_n(\rv) u_m(\rv) n_0(\rv) =0$ if $n \neq m$.  These eigenfunctions must be regarded as approximations to the matrix elements of the current density operator between the ground-state and the excited state in question, i.e.
$\uv_n(\rv) \simeq  \frac{[\jv]_{n0}(\rv)}{\omega_n n_0(\rv)}$, 
where $[\jv]_{n0}(\rv)\equiv \langle \Psi_n|\hat \jv(\rv)|\Psi_0\rangle$.  Even though this  
is only an approximation,  it is easy to verify that the sum rule $\sum_n \omega_n [j_\mu]_{n0}(\rv)[j_\nu]_{0n}(\rv') = m^{-2}M_{\mu\nu}(\rv,\rv')$ is satisfied by the approximate $[\jv]_{n0}(\rv)$.  In this sense our approximation preserves the total strength of the spectrum. {\it It is only for one-particle systems that the ``approximate" $[\jv]_{n0}(\rv)$
becomes exact}.  
Let us now illustrate the theory with two simple examples.
 
{\it Linear harmonic oscillator.}  For a harmonic oscillator of  frequency $\omega_0$, external potential $V_0(x)=m\omega_0^2x^2/2$, and equilibrium density $n_0(x) = \frac{e^{-x^2/l^2}}{\pi l}$, where $l \equiv (m\omega_0)^{-1/2}$,  the eigenvalue problem takes the form
\be\label{HOequation}
\frac{1}{4}u'''' -xu''' + (x^2 - 2)u''+ 3x u' - \frac{\omega^2-\omega_0^2}{\omega_0^2} u = 0\,,
\ee
where the prime denotes differentiation with respect to $x$.
Solving the eigenvalue problem with the boundary condition $n_0^{1/2}(x) u(x) \rightarrow 0$ for $|x| \rightarrow \infty$,
we obtain the exact excitation spectrum $\omega_n = \pm n\omega_0$,
where $n = 1, 2, ..$. The corresponding eigenfunctions are 
$u_n(x) \propto H_{n-1}(x)$.  These are indeed proportional to the matrix elements of the current density operator between the ground-state and the $n$-th excited state. 

 
 \begin{figure}
\begin{center}
\includegraphics[width=0.6\linewidth]{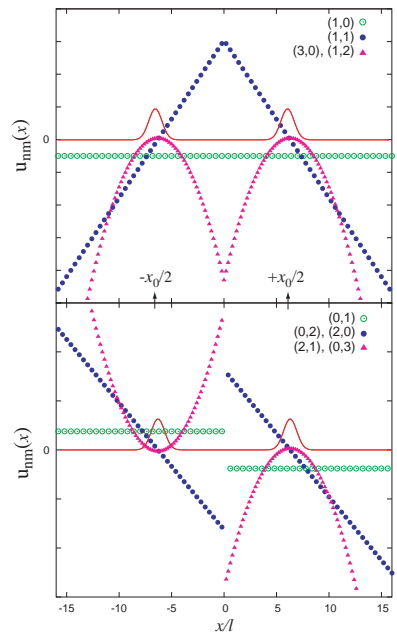}
\caption{(Color online) Unnormalized displacement fields for a few low-lying excitations of the two-electron system described in the text.  The solid line is the ground-state density. Analytically we find $u_{nm}(x) \propto H_{n+m-1}[2(x-x_0/2)]$  for $x \simeq x_0/2$, with parity $(-1)^{n-1}$ independent of $m$.  The large value  of the displacement field for $x\sim 0$ does not have a physical significance since the density is exponentially small in that region. \label{fig:one}}
\end{center}
\end{figure}

{\it Two-electron system.} Consider a system of two electrons
repelling each other with interaction potential
$\frac{e^2}{|x_1-x_2|}$ in a one-dimensional parabolic trap of  frequency $\omega_0$.
Due to the separation of center of mass and relative variables this model can easily be solved numerically, and even analytically in the limit of strong correlation.  We only focus on the strongly correlated limit ($\omega_0 \to 0$), since the non-interacting limit turns out to be exactly reproduced by our theory.   In the strongly correlated limit the two electrons become localized  near $x=\pm x_0/2$, where $x_0 = \left(2e^2/m\omega_0^2\right)^{1/3}$ is large (see Fig.~\ref{fig:one} for a plot of the density). The relative coordinate is a harmonic oscillator of frequency $\omega_0\sqrt{3}$  centered at $\pm x_0$.   The center of mass is  a harmonic oscillator with frequency $\omega_0$.  The exact eigenstates  are characterized by two non-negative integers $n$  (center of mass)  and $m$ (relative motion) and are denoted by $(n,m)$.   (0,0) is the ground-state.  The excitation energy associated with the state $(n,m)$ is  $E_{nm}=(n+m\sqrt{3})\omega_0$.  From the wave functions we calculate, without approximations,  the displacement field  $u_{nm}(x)$.  Some of the results are shown in Fig.~(\ref{fig:one}).   The displacement field of the $(1,0)$ excitation, which corresponds to a rigid translation of the center of mass, is uniform in space, while the displacement field of the $(0,1)$ excitation, which corresponds to the classical breathing mode, changes sign around the origin.  The $(1,0)$ and $(0,1)$ modes exhaust the classical phonon modes of a system of two localized particles.  The remaining excitations are fully quantum mechanical.  Examining Fig.~(\ref{fig:one}) one quickly realizes that all the excitations with a given value of $n+m$ and the same parity of $m$ produce the same displacement field, but have different energies.  
This is a feature of the exact solution that {\it cannot} be reproduced by any eigenvalue problem with a frequency-independent kernel.

\begin{table}
\begin{tabular}{|c|c|c|c|}
\hline 
$(n,m)$ & $\omega_{nm}^{exact}/\omega_0$  &$\omega_{nm}^{appr.}/\omega_0$ & $\bar\omega/\omega_0$\\
\hline 
(1,0) & 1.0 & 1.00 &1.00\\ 
\hline
(0,1) & 1.732 & 1.740 &1.732\\ 
\hline
(2,0) & 2.0 & 2.643 & 2.632 \\ 
(0,2)& 3.464 &~~~~&~~~~~\\  
\hline
(1,1) & 2.732 & 2.736&2.732\\ 
\hline
(3,0) & 3.0 & 3.950 & 3.942 \\ 
(1,2)& 4.464 &~~~~&~~~~~\\
\hline
(2,1) & 3.732 & 3.965 & 3.960 \\ 
(0,3)& 5.196 &~~~~&~~~~~\\
\hline
(4,0) & 4.0 & 5.224 & 5.217 \\ 
(2,2)& 5.464 &~~~~&~~~~~\\
(0,4)& 6.928 &~~~~&~~~~~\\
\hline 
\end{tabular}
\caption{Comparison between exact and calculated (appr.) excitation energies in the strongly correlated regime. The average frequency $\bar \omega$ of a group of excitations is calculated numerically from the sum rule discussed in the text. Analytically one finds $\bar \omega^2/\omega_0^2 = 2+3\sqrt{3}k +6k(k-1)(2-\sqrt{3})-(-1)^m (2-\sqrt{3})^k$, where $k\equiv n+m-1$: these exact values are indistinguishable, up to the third decimal digit, from the numerical results listed in the last column.}
\label{TableI}
\end{table}

Let us now see what our elastic equation of motion predicts for this system.  In Table I we present the energies of a few low-lying excitations  obtained from the numerical solution of Eq.~(\ref{eom.last}) in the strongly correlated regime.   We see that the energies of excitations such as $(1,0)$, $(0,1)$ and $(1,1)$, which do not ``share" their displacement field with other excitations,   are very well reproduced by our calculation within the accuracy of the numerical work.  On the other hand, groups of  excitations that share the same displacement field are replaced by a single excitation of average frequency, in such a way that the total spectral strength of the group is preserved.  Indeed, it can be proved that the average excitation frequency $\bar \omega$ that replaces the frequencies $\omega_l$ of the excitations in a given group is given by the sum rule 
$\bar \omega^2 = \sum_l f_l \omega_{l}^2$
where $f_l= \frac {2 m \left\vert \int d\rv  \jv_{0l}(\rv) \cdot \bar \uv (\rv)\right\vert^2}{\omega_{l}}$  is the ``oscillator strength" of the $l$-th excitation,  $\bar u(\rv)$ is the normalized solution of the eigenvalue problem with eigenvalue $\bar\omega^2$, and the sum runs over all the excitations in the group.  In the last column of Table I, we have checked that the sum rule  is quite well satisfied by the numerical solution of our two-electron model.

Before closing, we point out that the QCM formulation is  applicable directly to the Kohn-Sham system, in which case we do not need the exact ground-state density matrices, but only the ground-state Kohn-Sham orbitals and a reasonable approximation for the exchange-correlation field.  
%
The theory presented here is, in a very precise sense, the extension of the well-known collective approximation of the homogeneous electron gas to non-homogeneous systems   
and should therefore be useful in dealing with collective effects such as multi-particle excitations and the dipolar fluctuations that are responsible for van der Waals attraction.\cite{Langreth04,Dobson05}

This work was supported by DOE grant DE-FG02-05ER46203 (GV) and LDRD-PRD X9KU at LANL (JT) and by the Ikerbasque Foundation. GV gratefully acknowledges the kind hospitality of the ETSF in San Sebastian where this work was completed.


\begin{thebibliography}{100}

\bibitem{MonteCarlo}W. M. C. Foulkes, L. Mitas, R. J. Needs, and G. Rajagopal, Rev. Mod. Phys. {\bf 73}, 33 (2001).



\bibitem{grossbook}
{\it Time-Dependent Density Functional Theory}, Lecture Notes in Physics,
Vol. 706, edited by M.A.L. Marques, C.A. Ullrich, F. Nogueira,
A. Rubio, K. Burke, and E.K.U. Gross (Springer, Berlin, 2006).


\bibitem{Gross96} E.K.U. Gross,  J.F. Dobson and M. Petersilka,  in {\it Density
Functional Theory}, v. 181 of  {\it Topics in current chemistry}, 
ed. R.F. Nalewajski (Springer-Verlag, Berlin 1996).



\bibitem{Casida95}
M.E. Casida, in {\it Recent Advances in Density Functional Methods}, edited
by D.P. Chong (World Scientific, Singapore, 1995), p. 155.




\bibitem{Maitra04}
N. T. Maitra,  Fan Zhang,  R. J. Cave, and K.  Burke, J. Chem Phys. {\bf 120}, 5932 (2004).

\bibitem{Dobson98} J. F. Dobson and B. P. Dinte, in {\it Density Functional Theory}, Eds. J.
F. Dobson, G. Vignale and M. P. Das (Plenum, N.Y. 1998, 0-306-45834-9).

\bibitem{TerHaar}
D. Ter Haar {\em Introduction to the physics of many-body systems}
  (Interscience Publishers, London, 1958).

\bibitem{Madelung1927}
E. Madelung, {Z. Phys.} \textbf{{40}}, {322} ({1927}).

\bibitem{Bloch1933}
{F.}~{Bloch}, {Z. Phys.} \textbf{{81}}, {363} ({1933}).


\bibitem{GhoDeb1982}
{S. K.}~{Ghosh} {and} {B.~M.}~{Deb}, {Phys. Rep.} \textbf{{92}}, {1} ({1982}).

\bibitem{rg}
E. Runge  and E.K.U. Gross, Phys. Rev. Lett. {\bf 52}, 997 (1984).

\bibitem{vanLeeuwen99}  R. van Leeuwen, Phys. Rev. Lett. {\bf 82}, 3863 (1999).

\bibitem{Zaremba94} E. Zaremba and  H. C. Tso, Phys. Rev. B, {\bf 49}, 8147 (1994).

\bibitem{ConVig1999}
{S.}~{Conti} {and} {G.}~{Vignale},
{Phys. Rev. B} \textbf{{60}}, {7966} ({1999}).

\bibitem{TokPanPRB}
{I.}~{Tokatly} {and} {O.}~{Pankratov},
{Phys.\ Rev.\ B} \textbf{{60}}, {15550} ({1999}); {\it ibidem}  \textbf{{62}}, {2759} ({2000}).


\bibitem{DobLe}
{J.~F.}~{Dobson} {and} {H.~M.} {Le}, 
{J. Mol.\ Struct.: THEOCHEM} \textbf{{501--502}}, {327} ({2000}); {Phys. Rev. B} \textbf{{66}}, {075301} ({2002}).


\bibitem{tokatly}
I.V. Tokatly, Phys. Rev. B {\bf 71}, 165105 (2005); {\bf 75}, 125105 (2007).

\bibitem{tvt07}
J. Tao, G. Vignale, and I.V. Tokatly, Phys. Rev. B {\bf 76}, 195126 (2007).

\bibitem{Applications} {M.}~{Brewczyk}, {C.~W.}~{Clark}, {M.}~{Lewenstein}, {and} {K.}~{Rzazewski}, {Phys. Rev. Lett.} {\bf 80}, {1857} ({1998}); {P.}~{Hering}, {M.}~{Brewczyk}, {and} {C.}~{Cornaggia},
{Phys. Rev. Lett.} \textbf{{85}}, {2288} ({2000}); {Y.~E.}~{Kim} {and} {A.~L.}~{Zubarev},
{Phys. Rev. A} \textbf{{70}}, {033612} ({2004}).







\bibitem{gvbook}
G.F. Giuliani and G. Vignale, {\it Quantum Theory of the Electron Liquid},
(Cambridge University Press,  2005).

\bibitem{Langreth04}M. Dion, H. Rydberg, E. Schroder, D.C.  Langreth, and B.I. Lundqvist, Phys. Rev. Lett. {\bf 92}, 246401 (2004).

\bibitem{Dobson05}J. F. Dobson, Jun Wang, B. P. Dinte, K. McLennan and  H.
M. Le,  Int.  J. Quantum Chem. {\bf 101}, 579 (2005).


\end{thebibliography}
\end{document}